\documentclass[doublecol]{epl2} 
\usepackage{geometry}                
\usepackage{graphicx}
\usepackage{amssymb}
\usepackage{dsfont}
\usepackage{epstopdf}
\usepackage{color}
\usepackage{mathtools}
\usepackage{hyperref}

\definecolor{red}{rgb}{1,0,0}
\definecolor{blue}{rgb}{0,0,1}
\definecolor{black}{rgb}{0,0,0}

\newcommand{\p}{\partial}
\newcommand{\eq}[1]{\begin{align}#1\end{align}}

\newcommand{\ffrac}[2]{\mbox{$\frac{#1}{#2}$}}
\newcommand{\half}{\mbox{$\frac{1}{2}$}}
\newcommand{\OO}{\mathcal{O}}

\newcommand{\Tr}{\mbox{Tr}}
\newcommand{\D}{{\mathcal{D}}}

\usepackage{calrsfs}

\usepackage{scalerel,stackengine}
\stackMath
\newcommand\widecheck[1]{%
\savestack{\tmpbox}{\stretchto{%
  \scaleto{%
    \scalerel*[\widthof{\ensuremath{#1}}]{\kern-.6pt\bigwedge\kern-.6pt}%
    {\rule[-\textheight/2]{1ex}{\textheight}}
  }{\textheight}%
}{0.5ex}}%
\stackon[1pt]{#1}{\scalebox{-1}{\tmpbox}}%
}

\newcommand{\oR}{\overline{R}}
\newcommand{\og}{\overline{g}}
\newcommand{\Gt}{\tilde G}

\title{Glassy gravity}
\author{Eric De Giuli \inst{1} \and A. Zee \inst{2}}
\institute{
\inst{1}{Department of Physics, Ryerson University, M5B 2K3, Toronto, Canada} \\
\inst{2}{Kavli Institute for Theoretical Physics, University of California, Santa Barbara, CA 93106 - 4030, U.S.A.}
}
\abstract{
Euclidean quantum gravity is reconsidered, in the conformal mode approximation. Applying techniques from glass theory, we argue that the Euclidean partition function hides metastable states, which can be counted. This may reconcile conflicting results on the uniqueness of the de Sitter vacuum, and may be relevant to the cosmological constant problem. 
}
\begin{document}
\maketitle

Our current understanding of the universe is an unhappy marriage of quantum field theory and general relativity. This is perhaps most clearly seen in the cosmological constant (CC) problem: our na\"ive expectation of its value is off by some 120 orders of magnitude from its observed value \cite{Weinberg89,Nobbenhuis06,Porto10}. Many putative solutions to the CC problem, including anthropic rationalizations, involve the notion of a {\it multiverse}: that some quantities are dynamical in principle, but are fixed over the observable universe, or over observable timescales, and we observe just one of a multitude of possibilities \cite{Hawking84,Duff89,Weinberg87,Horava00}. This situation is equivalent to that encountered in the study of glass, for which an extensive number of degrees of freedom are fixed on experimental timescales \cite{Berthier11b}; particles only vibrate within the `cages' created by other particles. As a result, the system only explores a tiny fraction of phase space \cite{Monasson95,Franz95}.

In this Letter we present the hypothesis that Euclidean gravity, considered as a statistical field theory, is {\it glassy}. This would mean that there are exponentially many local minima of the action, which are not seen by the na\"ive partition function. These local minima are metastable states, also called `false vacua' in the context of QFT \cite{Coleman77,Coleman77a,Callan-Jr77}. Although absent from thermodynamics, these local minima will strongly affect the dynamics, and imply that the effective action strongly differs from the na\"ive one. 

Glassy systems typically break a symmetry of the Hamiltonian. It has been argued that de Sitter space is unstable to quantum fluctuations \cite{Mottola85,Anderson18,Krotov11,Mottola86,Polyakov08,Polyakov10,Anderson14}, and thus at most a false vacuum, but this latter result is debated. Calculations in Euclidean signature find a unique vacuum that can be analytically continued to Minkowski signature \cite{Marolf11}. It was argued in  \cite{Anderson14} that the unstable vacua are missed in the Euclidean computation because of the Euclidean regularity conditions. We suggest here that the unstable vacua are present as metastable states, just like in the partition function for a glass.

Techniques from glass theory can be used to count the metastable states, which give a configurational entropy, or {\it complexity}, $\Sigma$. The complexity vanishes in the liquid phase and becomes nonzero in the glassy phase \footnote{The complexity can vanish again at a lower temperature, the ideal glass (or Kauzmann) transition. In the ideal glass phase, the logarithm of the number of metastable states is subextensive.}.

Why should glassiness be present in quantum gravity? If space is unstable, then by definition it does not relax back to a simple vacuum; it retains some memory of the random perturbation, if in an encoded form. Each class of perturbations, presumably quantized at the Planck scale, defines a metastable state. Moreover, on a larger scale, a black hole, once formed, is difficult to eliminate. On timescales where the Hawking process can be ignored, black holes-- and possibly wormholes if they exist-- should create metastable states. 

If Euclidean quantum gravity is glassy, then the partition function $Z$ vastly {\it overcounts} states, since on observable timescales the universe is stuck in one sector of phase space. If we are stuck in an atypical vacuum, then $Z$ may not resemble at all the contribution $Z_{\text{local}}$ that we infer from local measurements. In principle, this discrepancy could soak up the putative huge contribution of the CC. 






To illustrate our ideas, we consider the conformal mode in 4D Euclidean quantum gravity. After integrating out massless matter fields, this becomes a physical mode through the trace anomaly \cite{Antoniadis07,Antoniadis92,Antoniadis92a,Antoniadis94,Mottola06}. We consider it here as a self-consistent gravitational theory, which could serve as a template for a more realistic computation including the graviton.
 We apply replica methods from the physics of disordered systems to compute the complexity of the Euclidean theory, in a variational approximation. We first consider $S^4$, the Euclidean continuation of de Sitter space. We find that although strict $S^4$ does not show metastable states, as soon as we add quenched fluctuations to the background metric, new states appear. In our model these new states are not stabilized to create true metastable states. However, this fact alone appears to reconcile earlier results \cite{Marolf11,Anderson14}. We find a large complexity of unstable states, suggesting that glassiness is present in quantum gravity.


\section{Metastable states} Metastable states appear in a statistical field theory when ergodicity is spontaneously broken. Although this is {\it a priori} a dynamical phenomenon, general methods to count the metastable states through thermodynamics were found by Monasson \cite{Monasson95}, whose construction we repeat here, and by Franz and Parisi \cite{Franz95}, whose related approach will be used in a later section. Consider an arbitrary statistical field theory in a field $\phi(x)$ with Hamiltonian $H$, to which we add an external pinning field $\psi(x)$, coupled through a term $U = h (\psi(x)-\phi(x))^2$. The free energy 
\eq{
F_{\phi}[\psi,h,\beta] = \frac{-1}{\beta} \log \int \D\phi \exp\left( -\beta H[\phi] - \half U \right)
}
will be small when the physical field is aligned with the pinning field (The field $\phi$ can have a vector index without any change in the arguments). By varying $\psi$, we can scan phase space to find metastable states. If these states remain when $h \to 0^+$, then ergodicity is spontaneously broken. In particular, consider the free energy of the probing field $\psi$ at inverse temperature $\beta m$, i.e.
\eq{
F_{\psi}(m,\beta) = \lim_{h\to 0} \frac{-1}{\beta m} \log \int\D\psi \exp\left(-m\beta F_{\phi}[\psi,h,\beta] \right)
}
The physical free energy is recovered for $m=1$, since in this case the pinning field is directly integrated out. We can decompose $F_\psi$ into `energetic' and `entropic' parts of the probing field via $E_\psi = \p(m F_\psi)/\p m|_{m=1}$ and $\Sigma = \beta \p F_\psi/\p m|_{m=1}$, respectively. Obviously $F_{\psi}|_{m=1} = E_\psi - \Sigma/\beta$. In simple models, it has been shown that the complexity $\Sigma$ computed in this way corresponds to the logarithm of the number of metastable states, computed explicitly \cite{Cavagna09}. For an extended discussion in the context of glasses, see \cite{Parisi20}. 

In quantum field theory, $\Sigma$ is the Shannon entropy of the pinning field.  To see this, define the trace over the pinning field as $\Tr \equiv \int \D\psi$ and note that its  distribution is $\rho[\psi] = e^{-\beta F_{\phi}}/Z$, where $Z = \Tr [e^{-\beta F_{\phi}}] = e^{-\beta F_{\psi}}|_{m=1}$. The Shannon entropy is
\eq{
-\Tr[\rho \log \rho] & = \left.\frac{-\p}{\p m}\Tr[\rho^m]\right|_{m=1}  \notag \\
& = -\left.\frac{\p}{\p m} \left[ \frac{\Tr[e^{-m\beta F_\phi}]}{Z^m} \right]\right|_{m=1} \notag \\
& = -\left.\frac{\p}{\p m} \left[ \frac{e^{-m\beta F_\psi}}{Z^m} \right]\right|_{m=1} = \beta \left.\frac{\p F_{\psi}}{\p m}\right|_{m=1} = \Sigma, \notag
}
as claimed.

It is clear that computing $F_\psi$ for arbitrary $m$ is intractable for general interacting theories. However, when $m$ is an integer, $\exp(-m\beta F_{\phi})$ can be represented by $m$ replicas of the original theory. The pinning field can then be integrated out to introduce a term $h/(2m) \sum_{a<b} \int dx (\phi_a(x)-\phi_b(x))^2$, where $a,b=1,\ldots,m$ label the replicas. This term attractively couples the replicas.  

The replicated theory can be treated by standard methods, and eventually analytically continued to $m=1$. In this analytical continuation rich, non-perturbative phenomena can appear \cite{Mezard87}. Importantly, since the pinning field is removed before the end of the computation, we do not add any new physical terms to the action. 


The replicas have a physical meaning as independent samples. In statistical field theory, the replicas should be considered as distinct systems, evolving with independent thermal fluctuations, but coupled together 
through the attractive interaction, which is eventually removed to probe for spontaneous breaking of ergodicity.
 In some cases these distinct replicas can be probed experimentally \cite{Seguin16}. In quantum field theory, the replicas have independent quantum fluctuations.

\section{Quantum gravity} Hitherto, the discussion has been for a general Euclidean QFT. We now specialize to gravity. Although a complete treatment of quantum gravity should include spin-2 fluctuations, such computations are at present intractable. We consider metrics conformally related to a reference metric $\og$, $g_{\mu\nu}(x) = e^{\sigma(x)} \og_{\mu\nu}(x)$. Although $\sigma$ is not dynamical in pure gravity, it becomes so upon integrating out all massless matter fields, through the trace anomaly \cite{Antoniadis92,Antoniadis94,Antoniadis07}. We hope that it captures the essential physics of glassiness in gravity. For simplicity we omit in what follows the kinetic trace anomaly term, although we have explicitly checked that our results are robust when it is included.

In the infrared regime the effective action for $\sigma$ is
\eq{ \label{S1}
S & = \half \int d^4 x \sqrt{\og} \left[ - \ffrac{6}{\kappa} e^{2 \sigma} \left[ (\overline{\nabla} \sigma)^2 + \ffrac{\oR}{6} \right] + 2 \lambda e^{4\sigma} \right].
}
The Euclidean partition function is $Z = \int \D\sigma \; e^{-S}$ and we call $F = -\log Z$ the free energy (absorbing a factor of $\beta$). As a result of the global Weyl symmetry it satisfies \cite{Antoniadis92,Antoniadis94,Antoniadis07}
\eq{ \label{weyl1}
F(\kappa e^{-2\omega},\lambda e^{4\omega}) = F(\kappa,\lambda),
}
indicating that the physical control parameter is $\lambda \kappa^2$. For simplicity in later calculations we will assume that this is small.  

Applying the Monasson construction to this action we add a replica index $a$ to $\sigma$, and add a term
\eq{
\frac{h}{2m} \int d^4 x \sqrt{\og} \sum_{a<b} (\sigma_a-\sigma_b)^2,
}
which couples the replicas. We will treat $S$ in the variational approximation \cite{Feynman98}, which is exact to one-loop and often an excellent approximation for disordered systems \cite{Bouchaud95,Mezard91,Mezard99}. We use a trial action that retains the quadratic part of $S$ and adds a translationally invariant self-energy operator $\Gamma_{ab}$ that couples the replicas, and a constant $k$, viz.,
\eq{ \label{S2}
S_V = \half \int d^4 x \sqrt{\og} \sum_a \sigma_a \left[ \sum_b (L_{ab} + \Gamma_{ab}) \sigma_b + k \right],
} 
where
\eq{
L_{ab} = \delta_{ab} \frac{6}{\kappa} \overline{\Box}  + \frac{h}{m} [\delta_{ab} - 1].
}
This corresponds to a propagator $G_{ab}=(L+\Gamma)^{-1}_{ab}$ and is the most general quadratic translationally invariant action. 

The variational free energy is
\eq{
F_V = -\log Z_V + \langle S - S_V \rangle,
}
where expectation values are taken with respect to $S_V$, for which $Z_V$ is the partition function. The true free energy $F$ satisfies $F \leq F_V$, thus the optimal approximation is obtained by extremizing $F_V$ with respect to the variational parameters $\Gamma_{ab}$ and $k$.

\section{4-sphere} We now let the reference metric be the Euclidean continuation of de Sitter space, i.e. the 4-sphere, for which $R^{\mu\nu} = \ffrac{1}{4} R g^{\mu\nu}$ and $R=12/r^2$ in terms of the radius $r$ of the sphere. To avoid complications with spherical harmonics, we approximate $\overline{\Box}$ by its flat space cousin $\nabla^2$ and work in momentum space with an IR cutoff\footnote{The Laplace-Beltrami operator has eigenvectors $Y_{kl}$ with eigenvalues $-l(l+3)/r^2$. Defining the IR cutoff by the smallest non-zero $l$, i.e, $l=1$, we set $-q_0^2=-4/r^2$.} $q_0=2/r$ for the non-constant modes, and a UV cutoff $\Lambda$. The latter parameterizes our ignorance regarding the unknown UV-completion of quantum gravity.

The full structure in replica space will be discussed below. Here we simply note that all diagonal elements of replica matrices are equal, i.e. $G_{aa}$ is independent of $a$, and similarly $G_X=\sum_{b} G_{ab}(q=0)$ is independent of $a$ \cite{Parisi20}. Up to an irrelevant constant we have
\eq{
\ffrac{1}{\Omega} \log Z_V & = \half \int_{q} (\log G(q))_{aa} + \ffrac{1}{8} k^2 m G_X
}
where $\int_{q} = \ffrac{1}{(2\pi)^4} \int_{q_0<q<\Lambda} d^4 q$ and $\Omega = 8 \pi^2 r^4/3$ is the space-time volume, and
\eq{
& \ffrac{1}{\Omega} \langle S - S_V \rangle = +\ffrac{1}{4} m k^2 G_X \notag \\
& \quad - m e^{2 G_0} e^{- k G_X} (\ffrac{3}{\kappa} G_2 + \ffrac{\oR}{2\kappa} ) \notag \\
& \quad + \lambda m  e^{8 G_0}  e^{-2 k G_X} - \ffrac{1}{8} k^2 G_X^2 J \notag \\
& \quad - \half m  \tilde\Omega + \ffrac{h}{2m} \sum_{a \neq b} \int_q G_{ab}(q)
}
Here $\tilde \Omega = \int_q 1$, $G_n = \int_{q} q^n G_{11}(q)$, and $J = \sum_{a,b} \Gamma_{ab}(q=0)$. Note that $G_{aa}$ is independent of $a$, thus the first replica is not singled out in $G_n$. 

The extremization over $\Gamma_{ab}$ and $k$ is sketched in the Supplementary Information (SI  \cite{SI}, which includes references to \cite{Parisi80,Parisi80a,Parisi80b,Mezard87}, \cite{Parisi20}, \cite{Rainone15,Rainone16}). The results are: (i) the variational approximation preserves the global Weyl symmetry if $G_X J = m$, which can be checked in the solution; (ii) as $m \to 1$ the solution is given by the {\it replica diagonal ansatz} where $G_{ab}(q) = \Gt(q) \delta_{ab}$; (iii) the complexity $\Sigma$ vanishes; 
 (iv) the Green's function $\tilde G(q)$ has the form
\eq{ \label{Gt1}
\Gt(q) = \frac{1}{\gamma q^2 + \mu},
}
When $\lambda \kappa^2 \ll 1$, as we assume, then $\mu \sim \Lambda^4/\log (1/(\lambda \kappa^2))$, $\gamma \sim -\Lambda^2/\log^2 (\lambda \kappa^2)$.

We have found that the complexity density vanishes: there is no spontaneous breaking of ergodicity. Does this depend on the replica-diagonal ansatz? First, we relax the replica-diagonal assumption, and consider a general replica-symmetric matrix $G_{ab} = \Gt \delta_{ab} + G (1-\delta_{ab})$. In this case, as shown in SI, all of the dependence on $G$ appears with a prefactor of $m-1$. Therefore as $m \to 1$, these terms add simple perturbative corrections to the above results and do not change the conclusions.

When metastable states are present, the replica symmetry must be broken. In the SI we show that this model does not have a replica-symmetry-breaking phase. There are therefore no metastable states in the pure $S^4$ model.


\section{4-sphere with background fluctuations} The fully symmetric space $S^4$ does not spontaneously break ergodicity: the complexity density vanishes when the pinning field is removed. Is this result robust with respect to changes in the reference metric?

We address this by adding to $S^4$ local fluctuations in the Ricci scalar. For simplicity, we only include the coupling of these fluctuations to the Einstein term, which is most relevant in the infrared. We thus consider
\eq{ \label{Sfluc}
S_{\text{fluc}}[\sigma] = S[\sigma] - \half \int_x \ffrac{1}{\kappa} e^{2\sigma(x)} \delta R(x)
}
We can ask, as above, whether ergodicity is spontaneously broken in $S_{\text{fluc}}$, at a fixed realization of $\delta R$, which acts as disorder. This requires that we compute $\Sigma$ from the disorder-averaged free energy
\eq{
\overline{F} = -\overline{ \log Z } = -\overline{\left.\frac{\p Z^n}{\p n}\right|_{n=0}} = -\left.\frac{\p \overline{Z^n}}{\p n}\right|_{n=0}, 
}
which indicates that we need to replicate the dynamical variables $n$ times. We can apply the Monasson construction to this object. We replicate $S_{\text{fluc}}$ from \eqref{Sfluc} $m$ times, add a pinning field, and then replicate $Z$ an additional $n$ times, so that there are $m\times n$ replicas of the $\sigma$ field. These $m \times n$ replicas are broken into $n$ groups of $m$ indistinguishable replicas \cite{Monasson95}. Within each group the replicas are coupled through the pinning field. Let us apply the variational procedure. 

The variational free energy is identical to the $S^4$ model, with 3 differences: first, there are now $nm$ replicas, where eventually $n\to 0$; second, the replica symmetry is already broken by the pinning field; and third, we add the term involving $\delta R$. Once averaged over the disorder, the latter will give
\eq{
\overline{ e^{\half \Sigma_a \int_x \frac{1}{\kappa} e^{2\sigma_a(x)} \delta R(x) } } = e^{\frac{1}{8\kappa^2} \Sigma_{a,b} \int_{x,x'} e^{2\sigma_a(x)} e^{2\sigma_b(x')} K(x,x')}, \notag
}
where we have assumed that $\delta R$ is Gaussian, with a kernel $K(x,x')=\langle \delta R(x) \delta R(x')\rangle$. We restrict ourselves to UV fluctuations, $K(x,x') = K \delta(x-x')$. The crucial technical point is that this term couples the replicas \cite{Mezard87}. Physically, this corresponds to the fact that the replicas share background metric fluctuations, as discussed more below. Such a coupling would arise from any distribution of the $\delta R$, and is a necessary but not sufficient condition to have replica symmetry breaking in this model \footnote{In the Discussion we outline how, in a theory going beyond the conformal mode, quenched fluctuations may not be necessary for replica symmetry breaking.}.

The free energy density is
\eq{ \label{fV2}
f_V & = -\half \int_{q} (\log G(q))_{aa}  \\
& \quad - nm e^{2 G_0} e^{-\tilde k} (\ffrac{3}{\kappa} G_2 + \ffrac{\oR}{2\kappa} ) + \lambda nm e^{8 G_0}  e^{-2\tilde k} \notag \\
& \quad + \ffrac{1}{8} k^2 G_X (m-G_X J) - \half nm \tilde\Omega  \notag \\
& \quad  - \ffrac{K}{8\kappa^2} \ffrac{1}{\Omega} \sum_{a,b} \int_{x} \langle  e^{2\sigma_a(x)} e^{2\sigma_b(x)} \rangle, \notag
}
where $\tilde k = k G_X$ and we have already dropped the infinitesimal pinning field \footnote{This field however has a crucial role in removing one degree of freedom in the $G_{ab}$ matrix. See \cite{Monasson95}.}. We have
\eq{
\langle  e^{2\sigma_a(x)} e^{2\sigma_b(x)} \rangle = e^{-2\tilde k} e^{2 \int_q (G_{aa}+G_{bb}+2G_{ab})}
}
The new term does not significantly alter the replica-symmetric solution from above, if $K \ll 1$. 
 The replica-diagonal part $\tilde G$ is as before, to $\OO(K)$. In the replica-symmetric ansatz, the off-diagonal part is
\eq{ \label{G1}
G(q) = \frac{K}{\kappa^2} e^{-2\tilde k} e^{4 G_0} \tilde G(q)^2 + \ldots
}
We now show, however, that this new term opens up the possibility for replica symmetry breaking, and hence ergodicity breaking.

The key is the nontrivial coupling of $G_{ab}(q)$ in the new term. Physically, $\tilde G(q) = G_{aa}(q)$ (no sum on $a$) is the propagator function as measured experimentally in one system. 
 Instead, $G_{ab}(q)$ for $a \neq b$ gives the correlations between two systems with the same fixed fluctuations in the background metric, but {\it independent} quantum fluctuations. Since the replicas have independent quantum fluctuations, this could be interpreted as regions of spacetime separated by a horizon. 
 


The result \eqref{G1} applies when any pair of replicas $a,b$ is equivalent. If spacetime is unstable, then we expect this to be too simple, because we can then have a replica $b$ that grew out of $a$ through instability, which may be inequivalent to replica $c$ that lives across a horizon. 

Thus we must probe for replica symmetry breaking. It was shown in \cite{Monasson95,Franz95} that it is sufficient to work in the neighborhood of $m=1$, where one can define a potential $V$ whose shape encodes the possibility of replica symmetry breaking. If $G_{ab}(q)$ for $a \neq b$ can only take two values, then after properly inserting this ansatz into $G_{ab}(q)$ we can create the Franz-Parisi potential $V[G_1(q)] = \p (f_V/(nm))/\p m|_{m=1}$ which is a functional of the propagator $G_1(q)$. $V$ measures the free energy cost to maintain the system with correlations $G_{ab}(q)=G_1(q)$, rather than $G(q)$ \cite{Franz95}. We choose $G_1$ to make the potential stationary, $G_1=G_1^*$, and then $V[G_1^*(q)] = \Sigma/\Omega$ is the complexity density \cite{Monasson95,Franz95}. The physical meaning of this last relation is that in the glassy phase, the system pays a free energy cost $V$ in exchange for creating a complexity density $\Sigma/\Omega$, since the total thermodynamic free energy $f=(f+V)-V$ does not see the metastable states. This assumes that the metastable states have a characteristic free energy.

The stationarity equation for $V$ always has a solution $G^*_1(q) = G(q)$ with $V=0$. When a second local minimum appears, there are metastable states \cite{Monasson95,Franz95}. It can be seen from \eqref{fV2} that the new term induced by background metric fluctuations always decreases the free energy, and is destabilizing. This is an artifact of our primitive method of adding background fluctuations and must be stabilized in a UV-complete quantum gravity theory. As a consequence, we find that $V$ gains a new stationary point, but which is a local maximum. We interpret $V$ evaluated at this point to be the density of {\it unstable} states. Thus replica symmetry breaking-- interpreted as the appearance of a new stationary point-- corresponds to instability of the background space, as argued for above.

We have
\eq{
& V[G_1] = \half \int_q \left[ \frac{(G_1(q)-G(q))(2G(q)-\tilde G(q))}{(\tilde G(q)-G(q))^2} \right. \notag \\
& \left. - \log \left( 1 - \frac{G_1(q)-G(q)}{\tilde G(q)-G(q)} \right) \right] + C \left[ e^{4 I} - e^{4 I_1} \right]
}
with $C = K(8\kappa^2)^{-1} e^{-2\tilde k} e^{4 G_0}$, $I = \int_q G(q)$, $I_1 = \int_q G_1(q)$. At a fixed value of the integral $I_1$, stationarity of $V$ yields a unique functional form of $G_1^*(q)$. The solution is then closed by setting $I_1 = \int_q G_1^*(q)$, leading to 
\eq{
\int_q \tilde G(q) - I_1 = \int_q \frac{\tilde G(q) - G(q)}{1 + 8 C [e^{4 I_1} - e^{4 I} ] (\tilde G(q) - G(q))} \notag
}
In the limit of small background fluctuations, $K \to 0$, this reduces to $I_1 - I = I [e^{4 I_1} - e^{4 I} ]$, which always has a nontrivial root with
\eq{
I_1 & = \begin{cases} I e^{-4 I} + \ldots & I \gg 1 \\ - \frac{1}{4} \log (4 I ) + \ldots &  I \ll 1 \end{cases}
} 
For $\lambda \kappa^2 \ll 1$, we find $I \sim K/\log^2(\lambda \kappa^2)$, using $C \sim K \Lambda^4/\log^4(\lambda \kappa^2)$. To leading order in $K$ we find that $G_1(q) = G(q) I_1/I$, i.e the correlations have the same spatial structure but differ in magnitude. After some calculations we find 
\eq{
V[G_1^*] \approx 2 C I_1^2/I, 
}
which is our main result. The complexity of unstable states behaves as $\Sigma = \Omega V~\sim~\Omega \Lambda^4 \log^2 K/\log^2(\lambda \kappa^2)$, which is approximately the number of Planck volumes in the universe, if $\Lambda \sim 1$ in Planck units. We emphasize that this quantity is not to be confused with the Bekenstein-Hawking entropy. Indeed, on a time-scale where the background metric fluctuations can be considered frozen, $V$ gives an estimate of {\it over-counting} in the na\"ive partition function, and thus acts in the opposite sense to traditional entropy.


\section{Discussion} We have applied methods from glass theory to count unstable modes in Euclidean quantum gravity, treated for simplicity with the conformal factor only. We have 
shown that once the background geometry is not fully symmetric, there are many unstable modes. These must be stabilized in a UV-complete theory of quantum gravity, thus realizing our claim that gravity is glassy. If the universe is stuck in a metastable state, then the partition function vastly overcounts states, and the effective gravitational parameters, such as the CC, may not resemble at all the bare ones. If the scenario described here survives more precise calculations, then this provides a partial explanation for why we may experience a tiny CC, uniquely suited to structure formation in the universe \cite{Weinberg89}. 



More generally, we hope to have convinced the reader that techniques from glass physics may be relevant to quantum gravity, and provide a new route to non-perturbative effects without adding any speculative physics beyond Einstein's theory. For example, replicas have recently been used as a technical tool in the computation of the path integral \cite{Engelhardt20} and spectral form factor in Jackiw-Teitelboim gravity \cite{Saad18,Saad19}. It was argued in these works that one needs to consider `replica wormholes', configurations that stitch together multiple replicas  \cite{Almheiri20,Almheiri20a,Almheiri20b,Engelhardt20}. Our claim is that importance of nontrivial coupling between replicas has a physical meaning, namely that gravity is glassy, which could be probed more precisely by computing the complexity $\Sigma$. 



Several important avenues for future work present themselves. First, one potential criticism of our approach is that the fluctuations in the background geometry have been introduced only through the Ricci scalar, and one may argue that these are either gauge artifacts, or would physically excite graviton modes, which we have neglected. A more convincing approach would thus be to consider fluctuations that cannot be removed locally, i.e. topological defects such as black holes or wormholes. These could be treated along the lines of \cite{De-Giuli20}. Alternatively, one could attempt to go beyond the conformal-mode approximation. If arbitrary geometries are considered in the path integral, rather than only fluctuations around a reference configuration, then the replicated path integral will have contributions from replica wormholes  \cite{Almheiri20,Almheiri20a,Almheiri20b,Engelhardt20}. Such configurations can lead to a nontrivial complexity $\Sigma$ without the need to introduce quenched background fluctuations.


Second, our treatment has been entirely in the Euclidean domain. It would be valuable to see how metastable states can be counted, explicitly, in the Minkowski domain. 

Finally, we have followed a logic, common in the physics of disordered systems, to extract essentially dynamical behavior from static computations. It would be worthwhile to treat the dynamics directly. It was argued recently, in a somewhat different context, that dynamics can be used to make a natural connection between classical disordered systems and quantum field theory \cite{Facoetti19}. 

\acknowledgments 
This work was initiated while A. Zee was visiting the \'Ecole Normale Sup\'erieure in Paris and CEA in Saclay. He thanks Henri Orland and \'Edouard Br\'ezin for hospitality and Jean-Philippe Bouchaud of CFM for financial support. We are grateful to John McGreevy, Emil Mottola, Pawel Mazur, and Edward Witten for comments on the manuscript, and to Francesco Zamponi for discussions about the replica method.

\bibliographystyle{eplbib}
\bibliography{../Gravity,../Glasses}

%

\end{document}






{\bf Supplementary Information.} Here we (1) sketch the extremization over $\Gamma_{ab}$ and $k$ in the $S^4$ model; and (2) show that the replica symmetry is not broken in the $S^4$ model. \\

{\bf Appendix 1. } The free energy density is
\eq{ \label{fV1}
f_V & = -\half \int_{q} (\log G(q))_{aa}  - m e^{2 G_0} e^{-\tilde k} (\ffrac{3}{\kappa} G_2 + \ffrac{\oR}{2\kappa} ) \notag \\
& \qquad + \lambda m e^{8 G_0}  e^{-2\tilde k} + \ffrac{1}{8} k^2 G_X (m-G_X J) - \half m \tilde\Omega  + \ffrac{h}{2m} \sum_{a \neq b} \int_q G_{ab}(q)
}
where $\tilde k = k G_X$, and $J = \sum_{a,b} \Gamma_{ab}(q=0) = \sum_{a,b} G^{-1}_{ab}(q=0) + h(m-1)$.

Let us check the Weyl symmetry in $F_V$. If we let $\lambda = \lambda' e^{4 \omega}, \kappa = \kappa' e^{-2\omega}$, then the $\omega$ factors in {\it both} of these terms can be absorbed into $k$ by a redefinition $k = k' + 2\omega/G_X$. This in turn induces a change in the free energy
\eq{
\delta F = (m-G_X J) \Omega \frac{\omega G_X k'  + \omega^2 }{2 G_X},
}
where we neglect terms that will vanish after imposing stationarity of $f_V$ with respect to the variational parameters. Thus the Weyl symmetry will hold for any $\omega$ if
\eq{ \label{weyl2}
G_X J = m,
} 
which we check below. 

Consider first the {\it replica diagonal ansatz} where $G_{ab}(q) = \Gt(q) \delta_{ab}$. In this case $G_X = \Gt(0)$ and 
\eq{ \label{weyl3}
G_X J -m = \Gt(0) m \left[ \frac{1}{\Gt(0)} + h (m-1) \right] -m = \Gt(0) m h (m-1).
}
Therefore as long as $\Gt(0)$ is $o(h^{-1})$ as $h\to 0$, the Weyl symmetry will remain intact when the pinning field is removed.

Stationarity with respect to $k$ gives
\eq{ \label{k}
0 &= \Gt(0) \left[  e^{2 G_0} e^{-\tilde k} (\ffrac{3}{\kappa} G_2 + \ffrac{\oR}{2\kappa}) - 2 \lambda e^{8 G_0} e^{-2\tilde k} - \ffrac{h (m-1)}{4} \tilde k   \right].
}
There are two branches. Either (i) $\Gt(0)=0$, and the Weyl symmetry holds even in the presence of the pinning field, or (ii) $\tilde k$ is a nontrivial root of this equation, and the Weyl symmetry is softly broken. We return to this below. We have
\eq{
f_V & = m \left[ -\half \int_{q} \log \Gt(q) -  e^{2 G_0} e^{-\tilde k} (\ffrac{3}{\kappa} G_2 + \ffrac{\oR}{2\kappa} ) + \lambda e^{8 G_0}  e^{-2\tilde k} - \ffrac{h(m-1)}{8} \tilde k^2 - \half \tilde\Omega \right]
}
The complexity density is
\eq{ \label{Sigma1A}
\Sigma/\Omega = 
\lim_{h\to 0} \p (f_V/m)/\p m|_{m=1} = 0 
}
Instead of extremizing $f_V$ with respect to $\Gamma_{ab}$, it is equivalent to extremize with respect to $G_{ab}$. In the replica diagonal ansatz we have only $\tilde G(q)$, thus we set 
\eq{
0 = \frac{1}{m} \frac{\delta f_V}{\delta \Gt(q)} & =  \ffrac{1}{m} \frac{\p f_V}{\p G_X} \delta_q - 2  e^{2 G_0} e^{-\tilde k} (\ffrac{3}{\kappa} G_2 + \ffrac{\oR}{2\kappa} + \ffrac{3}{2\kappa} q^2) + 8 \lambda e^{8 G_0} e^{-2\tilde k} - \frac{1}{2\Gt(q)},
}
where we now take $h = 0$. For both solution branches, $\p f_V/\p G_X=0$ from Eq.\eqref{k}, so that $\Gt(q)$ has the form
\eq{ \label{Gt1A}
\Gt(q) = \frac{1}{\gamma q^2 + \mu}.
}
with
\eq{
\mu & = - 4 e^{2 G_0} e^{-\tilde k} (\ffrac{3}{\kappa} G_2 + \ffrac{\oR}{2\kappa}) + 16 \lambda e^{8 G_0} e^{-2 \tilde k} \\
\gamma & = -\ffrac{6}{\kappa} e^{2 G_0} e^{-\tilde k}
}
when $h = 0$. Now we have that $\Gt(0)=1/\mu$, and we see that setting $\mu = \infty$ will lead to a divergent theory. We thus must take branch (ii), and choose $k$ to solve \eqref{k}. It follows that
\eq{
\mu = 8\lambda e^{8 G_0} e^{-2\tilde k}.
}
We need to compute $G_0 = \int_{q} \Gt(q)$ and $G_2 = \int_q q^2 \Gt(q)$. We have
\eq{
G_0 & = \frac{2\pi^2}{(2\pi)^4} \int_{q_0}^\Lambda dq \frac{q^3}{\gamma q^2 + \mu} 
}
To avoid the pole, we must have $\mu + \Lambda^2 \gamma >0$. We assume $\mu \gg -\Lambda^2 \gamma$ and show that this leads to a consistent solution. We have
\eq{
G_0 & = \frac{1}{8\pi^2} \frac{\Lambda^4}{4\mu} + \ldots \\
G_2 & = \frac{1}{8\pi^2} \frac{\Lambda^6}{6\mu} + \ldots
}
where we also use $\Lambda \gg q_0$. Then
\eq{
e^{-\tilde k} = e^{-6 G_0} \frac{3G_2}{2\lambda\kappa} + \ldots
}
and
\eq{
\mu = e^{-4 G_0} \frac{18 G_2^2}{\lambda\kappa^2} + \ldots
}
which is solved by
\eq{
\mu = -\frac{\Lambda^4}{24 \pi^2} \frac{1}{W(-(\lambda\kappa^2)^{1/3}/(3(2\pi)^{2/3}))},
}
where $W$ is Lambert's W function. The latter has two branches with either (i) $W(z) = z -z^2 + \ldots$ or (ii) $W(z) = \log(-z) - \log(-\log(-z)) + \ldots$ as $z \to 0$. To ensure that $\mu>0$, we require $\lambda\kappa^2<108\pi^{2}/e^3$. We have
\eq{
\frac{\mu}{-\gamma \Lambda^2} = \frac{2G_2}{\Lambda^2} + \ldots,
} 
which is large if $\mu \ll \Lambda^4$. This implies that we should take root (ii), and that we require $\lambda\kappa^2 \ll 1$ for this solution. \\

{\bf Appendix 2. Absence of replica symmetry breaking in the 4-sphere: } Here we show that no replica symmetry breaking solution exists, within the Parisi algebra that respects the global permutation symmetry of the problem. In this Appendix we use the notation $\hat{G}$ for the matrix $G_{ab}$.

First, consider the replica symmetric case $G_{ab} = \Gt \delta_{ab} + G (1-\delta_{ab})$. In this case
\eq{
\log \det \hat{G}(q) &  = m \log (\Gt(q)-G(q)) + \log \left(1 + m \frac{G(q)}{\Gt(q)-G(q)}\right) \\
G_X & = \Gt(q=0) + (m-1) G(q=0) \\
 \sum_{a,b} G^{-1}_{ab}(q) & =  \frac{m}{\Gt(q) + (m-1)G(q)} \\
 \sum_{a \neq b} \int_q G_{ab}(q) & = m(m-1) \int_q G(q)
}
and
\eq{
G_X J - m 
& = h (m-1) [\Gt(0)+(m-1)G(0)]
}
As $m \to 1$, $G$ is a regular perturbation, as claimed in the main text. 

Now consider the case of general $G_{ab}$. For each $q$, $G_{ab}$ is a $m \times m$ matrix, where eventually $m \to 1$. It is convenient to write $m=s+1$ and write $G$ as a block matrix with one distinguished replica, i.e.
\eq{
\hat{G} = \begin{bmatrix} \Gt & G_{1b} \\ G_{a1} & \hat{G}^{(s)} \end{bmatrix} 
}
The lower-right $s \times s$ part belongs to the Parisi hierarchical algebra \cite{Parisi80,*Parisi80a,*Parisi80b,Mezard87}. In the analytic continuation to $s \to 0$, this matrix is specified by its diagonal part $\Gt$, which is constant between replicas, and a function $G(x)$ on the interval $s < x < 1$, which encodes the correlations between the replicas \cite{Parisi20}. This $x$ is not to be confused with a spatial index: it actually has an interpretation in terms of distance between states \cite{Mezard87}. For a general model, the function $G(x)$ can take several different forms: (i) when metastable states are absent, $G(x)=G$ is constant, and the construction reduces to the replica-symmetric form considered in the main text; (ii) a non-trivial $G(x)$ can be piecewise-constant, with two segments. This is the one-step replica-symmetry breaking regime; (iii) the function $G(x)$ can have a region of $x$ in which it varies continuously. This is the infinite step replica-symmetry breaking regime, and describes marginally-stable phases with a fractal energy landscape. In principle this is the limit of a piecewise function with infinitely many segments, but phases with $2 \leq k < \infty$ segments are extremely rare.

In $\hat{G}$ the vectorial parts $G_{1b}$ and $G_{a1}$ cannot depend on $b$ and $a$ \cite{Parisi20}, similar to the so-called state-following procedure \cite{Rainone15,*Rainone16}. We set them equal to $G_r = \ffrac{1}{1-s}\int_s^1 dx G(x)$ to ensure that the reference replica is identical to the others. Using results from \cite{Rainone16,Parisi20} we have, to $\OO(s)$ as $s \to 0$,
\eq{
G_X & = \Gt(q=0) + s G_r(q=0)\\
\log \det \hat{G}(q) & = \log \det \hat{G}^{(s)}(q) + \log \Gt(q) - \frac{s G_r(q)^2}{\Gt(q)(\Gt(q) - G_r(q))} \\
 \log \det \hat{G}^{(s)}(q)  & = s \log (\Gt(q) - G_r(q)) + s \frac{G(x=0,q)}{\Gt(q)-G_r(q)} - s \int_s^1 \frac{dy}{y^2} \log \left[ 1 - \frac{[G](y,q)}{G(q)-G_r(q)} \right] \\
 \sum_{a,b} G^{-1}_{ab} & =  \frac{1}{\Gt} + \frac{s (\Gt - G_r)}{\Gt^2} \\
 \sum_{a \neq b} \int_q G_{ab}(q) & = s \int_q G_r(q) 
 }
where 
\eq{
[G](x,q) = x G(x,q) - \int_s^x dy G(y,q)
}
Note that $\delta [G](y,q)/\delta G(x,q) = x \delta(x-y) - \Theta(y-x)$ and $\p [G](x,q)/\p x = x \p G(x,q)/\p x$.

Let us first extremize with respect to $G(x,q)$ for $q>0$. Since the result depends only on one fixed $q$, we suppress this dependence. We have
\eq{
0 = \frac{\delta f_V}{\delta G(x)} & = \frac{s/2}{\Gt-G_r} + \ffrac{s}{2} \int_s^1 \frac{dy}{y^2} \left[ \frac{\Gt - G_r - [G](y)}{\Gt-G_r} \right]^{-1} \left[ -\frac{x \delta(x-y)-\Theta(y-x)}{\Gt-G_r}- \frac{[G](y)}{(\Gt-G_r)^2} \right] \\
& + \frac{s G(x=0)}{(\Gt - G_r)^2}  + \frac{s G_r}{\Gt(\Gt - G_r)} + \frac{\half s G_r^2}{\Gt(\Gt - G_r)^2}  + \half h s  + \OO(s^2)
}
which gives
\eq{
0 & = \frac{1}{\Gt-G_r} - \frac{1}{x} \frac{1}{\Gt - G_r - [G](x)}  + \int_x^1 \frac{dy}{y^2} \frac{1}{\Gt - G_r - [G](y)} \notag \\
& - \frac{1}{\Gt-G_r} \int_s^1 \frac{dy}{y^2} \frac{[G](y)}{\Gt-G_r - [G](y)} + \frac{2 G(x=0)}{(\Gt - G_r)^2} + \frac{2G_r}{\Gt(\Gt - G_r)} + \frac{G_r^2}{\Gt(\Gt - G_r)^2}  + h + \OO(s),
}
and, by differentiation,
\eq{
0 & = - \frac{\p_x G}{(\Gt - G_r - [G](x))^2},
}
so that $\p_x G(x)=0$. This implies that an infinite replica-symmetry-breaking solution is not possible. Note that this result does not depend on the form of the interactions, as long as they are local. 

This implies that $G(x)$ is piecewise-linear. There is no requirement that $G(x)$ be continuous, and indeed this function is typically found to have several segments. Consider the so-called 1RSB regime where
\eq{
G(x) = \begin{cases} g_0 & \mbox{for } x<x_0 \\ g_1 & \mbox{for } x>x_0 \end{cases}
}
In this case $[G](x)=x_0(g_1-g_0)$ for $x>x_0$ and is 0 otherwise, and $G_r = x_0 g_0 + (1-x_0) g_1$. The $G$ dependent part of the action is
\eq{
f_G = s\half \int_q \left[ -\log(\Gt-G_r) - \frac{g_0}{\Gt-G_r} + (x_0^{-1}-1) \log \left[ 1 - \frac{x_0(G_r-g_0)}{(1-x_0)(\Gt-G_r)} \right] + \frac{G_r^2}{\Gt(\Gt-G_r)} + h G_r \right]
}
Instead of extremizing $f_V$ with respect to $g_0,g_1$, and $x_0$, we can extremize with respect to $g_0$, $G_r$, and $x_0$. The equations are degenerate and give
\eq{
G_r = g_0 = -\frac{h \Gt^2}{1-h \Gt},
}
which implies either $x_0=1$ or $g_1=g_0$. In either case we recover the replica symmetric solution. Similar arguments apply to kRSB with $k>1$. Therefore no replica symmetry breaking is possible.

\bibliographystyle{abbrv}
\bibliography{../Gravity,../Glasses}